\documentclass[twocolumn]{aastex63}
\usepackage{float}
\usepackage{graphicx}
\usepackage{amsmath}
\usepackage{natbib}
\usepackage{color}
\usepackage{verbatim}
\usepackage[utf8]{inputenc}
\usepackage{mathtools}
\usepackage{amssymb}
\usepackage{textcomp}
\usepackage{enumitem}
\usepackage{braket}
\usepackage{totcount}
\usepackage{footnote}

\renewcommand{\thetable}{\arabic{table}}

\newtotcounter{citnum} 
\def\oldbibitem{} \let\oldbibitem=\bibitem
\def\bibitem{\stepcounter{citnum}\oldbibitem}

\begin{document}
\title{Origins of the Evil Eye: M64's Stellar Halo Reveals the Recent Accretion of an SMC-mass Satellite}

\author[0000-0003-2599-7524]{Adam Smercina}
\affiliation{Department of Astronomy, University of Washington, Box 351580, Seattle, WA 98195-1580, USA}
\author[0000-0002-5564-9873]{Eric F. Bell}
\affiliation{Department of Astronomy, University of Michigan, 1085 S. University Ave, Ann Arbor, MI 48109-1107, USA}
\author[0000-0003-0511-0228]{Paul A. Price}
\affiliation{Department of Astrophysical Sciences, Princeton University, Princeton, NJ 08544, USA}
\author[0000-0001-6380-010X]{Jeremy Bailin}
\affiliation{Department of Physics and Astronomy, University of Alabama, Box 870324, Tuscaloosa, AL 35487-0324, USA}
\author[0000-0002-1264-2006]{Julianne J. Dalcanton}
\affiliation{Center for Computational Astrophysics, Flatiron Institute, 162 Fifth Avenue, New York, NY 10010, USA}
\affiliation{Department of Astronomy, University of Washington, Box 351580, Seattle, WA 98195-1580, USA}
\author[0000-0001-6982-4081]{Roelof S. de Jong}
\affiliation{Leibniz-Institut f\"{u}r Astrophysik Potsdam (AIP), An der Sternwarte 16, D-14482 Potsdam, Germany}
\author[0000-0001-9269-8167]{Richard D'Souza}
\affiliation{Department of Astronomy, University of Michigan, 1085 S. University Ave, Ann Arbor, MI 48109-1107, USA}
\affiliation{Vatican Observatory, Specola Vaticana, V-00120, Vatican City State} 
\author[0000-0003-2294-4187]{Katya Gozman}
\affiliation{Department of Astronomy, University of Michigan, 1085 S. University Ave, Ann Arbor, MI 48109-1107, USA}
\author[0000-0002-2502-0070]{In Sung Jang}
\affiliation{Department of Astronomy and Astrophysics, University of Chicago, Chicago, IL 60637, USA}
\author[0000-0003-2325-9616]{Antonela Monachesi}
\affiliation{Instituto de Investigaci\'on Multidisciplinar en Ciencia y Tecnolog\'ia, Universidad de La Serena, Ra\'ul Bitr\'an 1305, La Serena, Chile}
\affiliation{Departamento de Astronom\'ia, Universidad de La Serena, Av. Juan Cisternas 1200 N, La Serena, Chile}
\author[0000-0002-1793-3689]{David Nidever}
\affiliation{Department of Physics, Montana State University, P.O. Box 173840, Bozeman, MT 59717-3840}
\author[0000-0002-0558-0521]{Colin T. Slater}
\affiliation{Department of Astronomy, University of Washington, Box 351580, Seattle, WA 98195-1580, USA}

\email{asmerci@uw.edu}

\begin{abstract}
M64, often called the ``Evil Eye'' galaxy, is unique among local galaxies. Beyond its dramatic, dusty nucleus, it also hosts an outer gas disk that counter-rotates relative to its stars. The mass of this outer disk is comparable to the gas content of the Small Magellanic Cloud (SMC), prompting the idea that it was likely accreted in a recent minor merger. Yet, detailed follow-up studies of M64's outer disk have shown no evidence of such an event, leading to other interpretations, such as a ``fly-by'' interaction with the distant diffuse satellite Coma P. We present Subaru Hyper Suprime-Cam observations of M64's stellar halo, which resolve its stellar populations and reveal a spectacular radial shell feature, oriented $\sim$30\textdegree\ relative to the major axis and along the rotation axis of the outer gas disk. The shell is $\sim$45\,kpc southeast of M64, while a similar but more diffuse plume to the northwest extends to $>$100\,kpc. We estimate a stellar mass and metallicity for the southern shell of $M_{\star}\,{=}\,1.80\,{\pm}\,0.54{\times}10^8\,M_{\odot}$\ and [M/H]\,$=$\,$-$1.0, respectively, and a similar mass of $1.42\,{\pm}\,0.71{\times}10^8\,M_{\odot}$\ for the northern plume. Taking into account the accreted material in M64's inner disk, we estimate a total stellar mass for the progenitor satellite of $M_{\rm \star,prog}\,{\simeq}\,5{\times}10^8\,M_{\odot}$. These results suggest that M64 is in the final stages of a minor merger with a gas-rich satellite strikingly similar to the SMC, in which M64's accreted counter-rotating gas originated, and which is responsible for the formation of its dusty inner star-forming disk. 
\\
\end{abstract}

\section{Introduction}
\label{sec:intro}

\begin{figure*}[t]
    \centering
    \includegraphics[width={0.9\linewidth}]{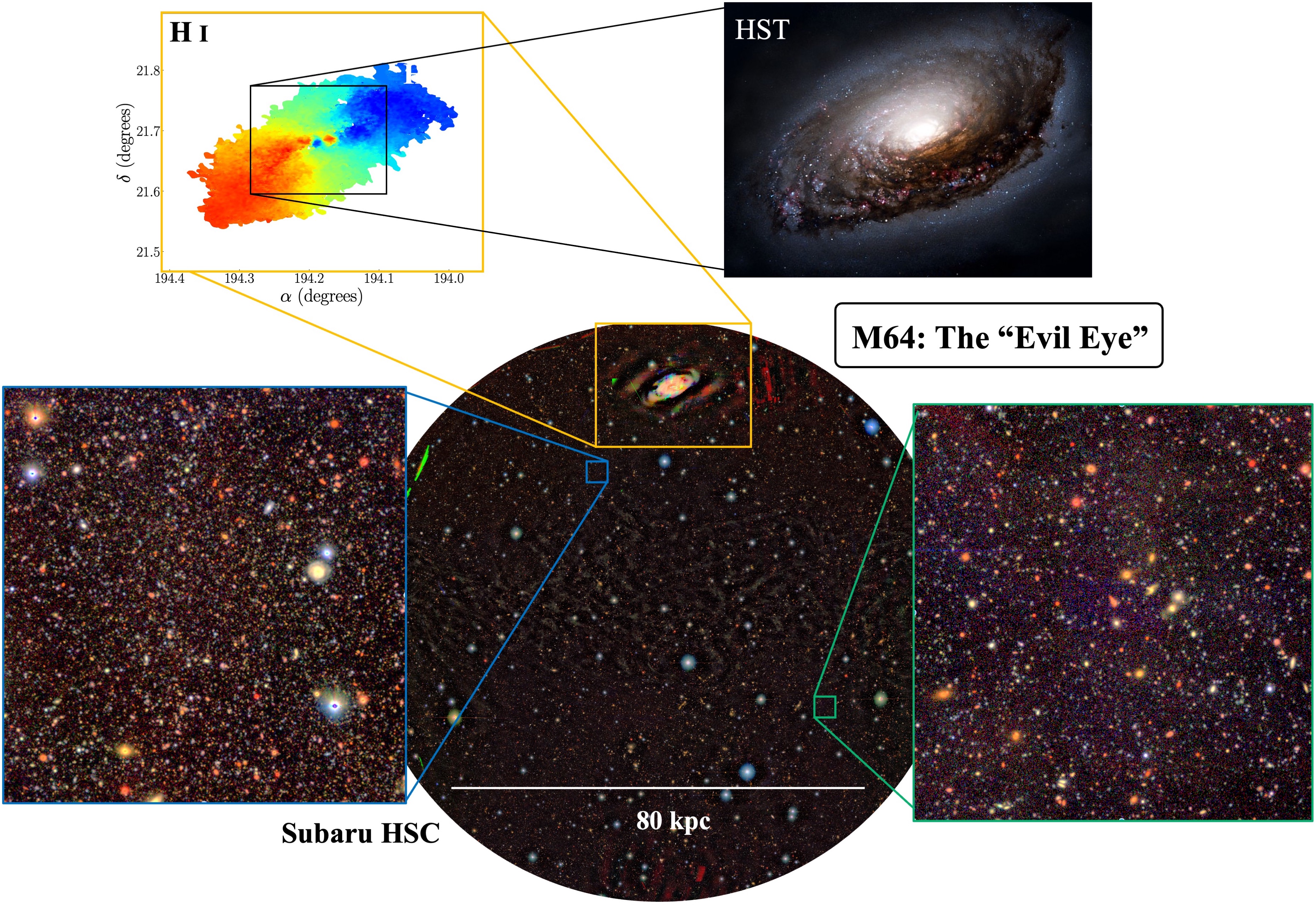}
    \caption{\textit{Bottom center}: $gri$\ color mosaic of observations taken with Subaru HSC in the deepest of the three fields, Field 3. M64 is visible at the top of the field, which is $\sim$115\,kpc across at its widest, with much of the diffuse brightness associated with the disk subtracted off as part of the reduction. Obfuscating Galactic cirrus is visible as diffuse filamentary features over much of the field, the color of which complicates inferences of M64's stellar halo from integrated light. The observations are displayed with a logarithmic stretch. \textit{Bottom insets}: 3\arcmin\,$\times$\,3\arcmin\ zoomed regions in M64's halo, displayed at the same stretch as the larger image, showing examples of areas dominated by resolved stars at the distance of M64 (left, blue) and background galaxies (right, green). \textit{Top right}: Color HST image of the inner region of M64 from the \href{https://hubblesite.org/contents/media/images/2004/04/1447-Image.html?keyword=M64&filterUUID=6b40edb4-2a47-4f89-8047-2fe9359344f3}{Hubble Heritage project}. M64's dusty inner disk, from which it derives its nickname the ``Evil Eye'', is prominent. \textit{Top left}: \textsc{H\,i} Moment 1 velocity map, taken from the THINGS survey \citep{walter2008}. M64's two gas disks are visible, with the larger outer disk rotating counter to the inner disk and the bulk of M64's stars.}
    \label{fig:m64-mosaic}
\end{figure*}

Merging has long been thought to be an important mode of evolution for galaxies like the Milky Way (MW) \citep[e.g.,][]{toomre&toomre1972,barnes&hernquist1996}, complemented by ``secular'' evolution of the disk \citep[e.g.,][]{kormendy&kennicutt2004} and direct gas accretion from the intergalactic medium (IGM) \citep[e.g.,][]{tumlinson2017}. Disentangling these processes to understand which present-day galaxy properties are a result of the merger history is a challenging task, as the impact of mergers (especially minor ones) on a MW-mass galaxy's disk can be subtle. The nearby galaxy M64 (also NGC 4826) is a prime example. Long nicknamed the ``Evil Eye'' galaxy due to the presence of an unusual dusty inner gas disk (Figure \ref{fig:m64-mosaic}, top left), M64 is unique in the Local Volume in hosting an outer \textsc{h\,i} disk that counter-rotates \citep{braun1992} with respect to its stellar disk (see Figure \ref{fig:m64-mosaic}, top right). The collision between these two disks, with its high relative velocity, is thought to have produced an enhanced star formation rate at their interface, which is likely responsible for M64's infamous visible dust lanes \citep{braun1992}. This idea is supported by much more recent, larger studies of galaxies with dual counter-rotating gas disks \citep{cao2022}. 

Since its discovery, the prevailing assumption has been that M64's counter-rotating gas disk was accreted during a recent merger or interaction \citep[e.g.,][]{rubin1994}. Yet, puzzlingly, M64's outer disk shows little sign of a recent interaction in either stellar kinematics \citep{rix1995} or deep ground-based imaging \citep[][see also Figure \ref{fig:m64-mosaic}]{watkins2016}. This prompted the early idea that M64 may have accreted its counter-rotating gas directly from the IGM \citep{rix1995}. However, more recent speculation has suggested a ``fly-by'' interaction within the last $\sim$1\,Gyr with the diffuse, gas-rich dwarf galaxy Coma P may be responsible \citep{brunker2019}. There has been evidence for M64 having a stellar halo, as first uncovered by \cite{kang2020}, who found a distinct metal-poor stellar halo population in a single Hubble Space Telescope field along M64's major axis. Yet, with only a single field it is completely ambiguous whether this population was accreted along with M64's counter-rotating gas, during, for example, an interaction with Coma P or an as-yet unknown progenitor or if it was deposited by more ancient accretions. 

The approach of resolving bright red giant branch (RGB) stars in nearby galaxies with wide-field ground-based imagers is ideally suited to this problem \citep[e.g.,][and others]{bailin2011,ibata2014,okamoto2015,smercina2020}. The debris from any interaction between M64 and a galaxy massive enough to account for its counter-rotating gas, either as a flyby or a full disruption, should be easily visible as an overdensity of point sources with colors and magnitudes consistent with a population of RGB stars at the distance of M64. 

In this Letter, we present Subaru Hyper Suprime-Cam (HSC) observations of M64's stellar halo, resolving its stellar populations in three pointings and reaching radii of $\sim$115\,kpc. Throughout this Letter, we assume a distance to M64 of 4.4\,Mpc \citep{tully2013} and a total stellar mass of $M_{\rm\star,Gal}\,{=}\,2.8{\times}10^{10}\ M_{\odot}$\ \citep{querejeta2015}. We find evidence for a spectacular shell feature, and other tidal structures indicative of an ongoing, late-stage radial merger. Benchmarking against existing Hubble Space Telescope (HST) photometry, we estimate the stellar mass of the progenitor galaxy to be $M_{\star}\,{\simeq}\,5{\times}10^8\,M_{\odot}$, with a metallicity of [M/H]\,$\simeq$\,$-$1 --- very similar to the mass and metallicity of the Small Magellanic Cloud (SMC). The mass of M64's counter-rotating outer gas disk, $M_{\rm gas,CR}\,{\simeq}\,5{\times}10^8\ M_{\odot}$\ \citep{walter2008}, is also comparable to the gas mass of the SMC \citep{dickey2000}, suggesting that the likeliest origin of M64's unique counter-rotating disk was a recent merger with a gas-rich satellite very similar to the SMC. 

\section{Observations, Reduction, and Artificial Star Tests}
\label{sec:obs}

Subaru HSC observations were carried out in Subaru HSC's Queue Mode during the 2019A semester, through the Gemini--Subaru exchange program (PI: Smercina, 2019A-0283). Our survey of M64's stellar halo consists of three 1\fdg5 HSC pointings: two northern fields (Fields 1 \&\ 2) with imaging in $g$\ and $r$\ bands, and a third southern field (Field 3) with imaging in $g$, $r$, and $i$\ bands, for a total of 5.2\,hr of exposure time. These three fields cover an effective area equivalent to a circle of radius $\sim$100\,kpc, centered on M64.

The data were reduced using the most up-to-date version of the HSC optical imaging pipeline \citep{bosch2018}, which produces final magnitudes in the Sloan Digital Sky Survey (SDSS) filter system. Sources were detected separately in each available filter and photometry was performed on the coadded stack of all available images. For each source, a clipped mean of deblended pixels is measured in an annulus that spans from 7 to 15 times the 68\% enclosed light radius of the point spread function (PSF). This source-specific background measurement, normalized to the effective aperture area, is then subtracted from the source fluxes. All magnitudes were then corrected for Galactic extinction using the maps of \cite{schlafly&finkbeiner2011}.\footnote{Using the Python package \href{https://github.com/kbarbary/sfdmap}{\texttt{sfdmap}}.} Seeing conditions were excellent, yielding measured FWHM between 0\farcs4--0\farcs75 for the brightest sources in all three fields.

While the initial survey design provided three-filter coverage to equal depth across all three fields, poor weather on Maunakea in the 2019A semester substantially reduced the available observing time in the required good/excellent seeing conditions. The survey was, therefore, only partially completed, and Subaru does not carry over programs do subsequent semesters. Given this, throughout the paper we base most quantitative inferences of M64's halo properties on our completed Field 3 observations due to the immense benefit of color--color information in selecting stellar sources from its three-band photometry \citep[see][for further details]{smercina2017,smercina2020}. For Field 3, we conducted a suite of artificial star tests (ASTs) to characterize the photometric quality and completeness of the data. Artificial stars were drawn uniformly from $g{-}i$\ vs.\ $i$\ color--magnitude space, with $22.6\,{<}\,i\,{<}\,28.6$\ and $0\,{<}\,g{-}i\,{<}\,3.5$. $r$-band magnitudes were then calculated using the $g{-}r$/$r{-}i$\ stellar locus (see \S\,\ref{sec:results}). Artificial stars were then inserted uniformly across the field in 10 separate runs, with 87,800 stars per run, for a total of 878,000 artificial stars. Following \cite{smercina2020}, we consider an artificial star ``recovered'' by the pipeline if the detected source is within 0\farcs3 of the input position and within 1\,mag of the input magnitude in each filter. From these ASTs, we estimate general 50\% source completeness limits of $g_0\,{=}\,27.44$, $r_0\,{=}\,26.69$, and $i_0\,{=}\,25.91$\ for Field 3. 

\begin{figure*}[t]
    \begin{minipage}{\linewidth}
        \centering
        \includegraphics[width={\linewidth}]{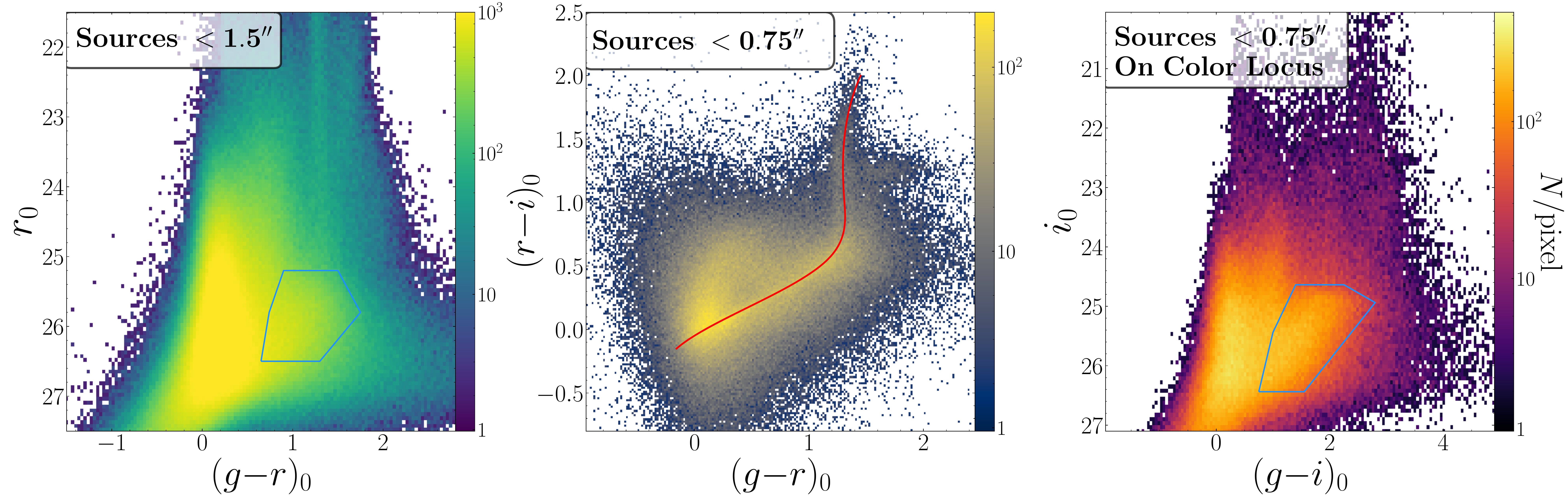}
    \end{minipage}
    \par\vspace{\baselineskip}
    \begin{minipage}{\linewidth}
        \centering
        \includegraphics[width={0.8\linewidth}]{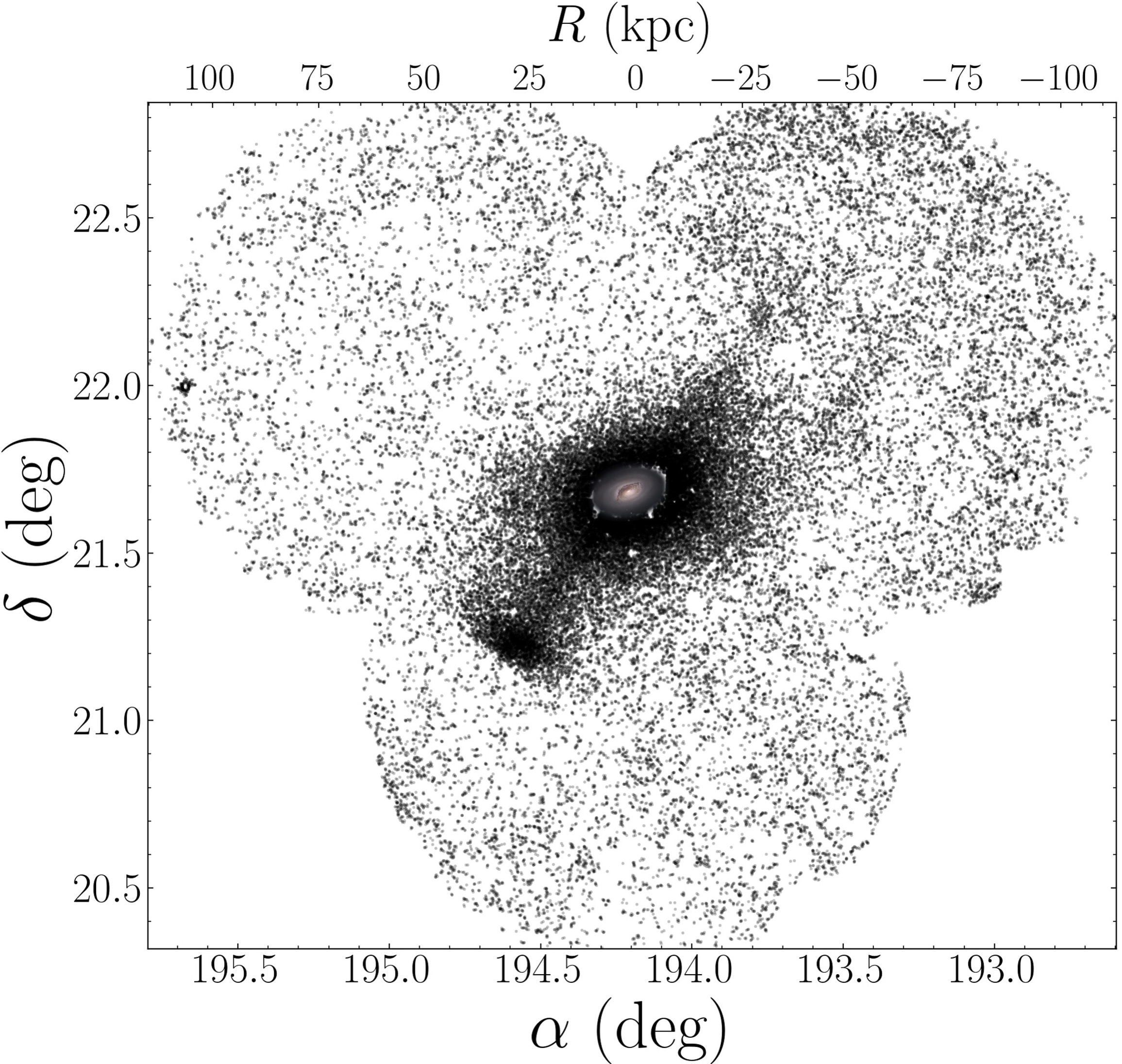}
    \end{minipage}
    \caption{\textit{Top left}: Hess CMD in $g$\ and $r$\ of all sources $<$1\farcs5 in size in Fields 1, 2, \&\ 3. The RGB selection region is shown in blue. The considerable contamination from background galaxies and MW foreground stars is clearly visible blueward and above the RGB, respectively. \textit{Top center}: Color--color diagram of all sources in the southernmost Field 3 (only field with three filter coverage) $<$0\farcs75 in size. The color locus for stars is shown in red. \textit{Top right}: Hess CMD of the highest-quality stellar candidates in the southern Field 3: sources $<$0\farcs75 in size and on the color locus, within photometric uncertainties. \textit{Bottom}: Spatial map of sources $<$1\farcs5 in size, witin the $g{-}r$\ vs.\ $r$\ RGB selection region. Modest nearest-neighbor filtering has been applied to remove contamination, highlighting the higher-surface brightness structures in M64's halo. The interior of M64 has been filled in with a ground-based image from the Jacobus Kapteyn Telescope (Nik Szymanek, Isaac Newton Group of Telescopes \href{https://www.ing.iac.es//PR/science/galaxies.html}{archive}). A spectacular radial shell feature is visible in the southern halo, with a more diffuse opposing plume extending to the north.}
    \label{fig:m64-wide}
\end{figure*}

\begin{figure*}[t]
    \centering
    \includegraphics[width={\linewidth}]{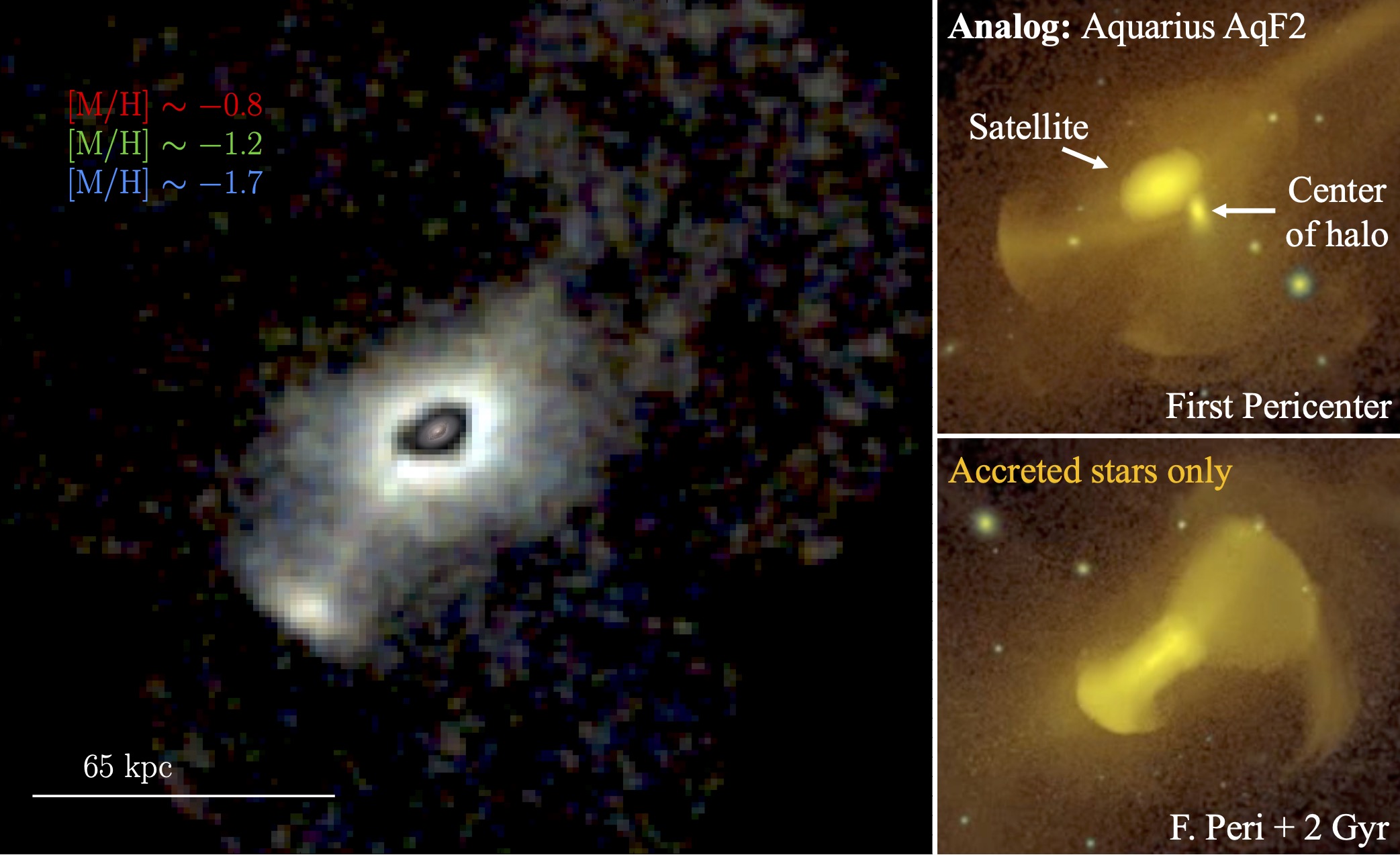}
    \caption{\textit{Left}: Map of loosely selected RGB stars in M64's halo (see also Figure \ref{fig:m64-wide}, bottom), with intensity mapped to stellar density, where each `channel' represents stars in three bins of metallicity: [M/H]\,${\sim}\,{-}$0.8 (red), [M/H]\,${\sim}\,{-}$1.2 (green), and [M/H]\,${\sim}\,{-}$1.7 (blue). Each channel was clipped to two sources per pixel and then smoothed with a Gaussian filter of width $\sim$1\,kpc. The interior of M64 has been filled in with a ground-based image from the Jacobus Kapteyn Telescope.     \textit{Right}: Two snapshots from the Aquarius AqF2 simulation presented in \cite{cooper2011}, showing the distribution of accreted stellar material\,{\color{blue}$^{\rm a}$} around a central galaxy of similar mass to M64 during a minor merger with a satellite on a radial orbit. Each snapshot is approximately 80\,kpc\,$\times$\,70\,kpc. The top panel shows the satellite at first pericenter and the lower panel at a point $\sim$2\,Gyr later. While only qualitative, the lower panel closely resembles the distribution of accreted stars in M64's halo, including its southern shell and northern plume.\\
    {\color{blue}$^{\rm a}$} \footnotesize{Note that these images do \textit{not} show the stars formed in the halo of the central galaxy, which is why, as noted by \cite{cooper2011}, the material at the center of the halo appears fainter than the satellite.}}
    \label{fig:m64-channel}
\end{figure*}

\section{Detecting M64's Halo: A Spectacular Radial Merger Remnant}
\label{sec:results}

In ground-based resolved star surveys at the photometric depths probed here, the separation of stars at the same line-of-sight distance as the host galaxy and more distant background galaxies, many of which appear as point sources in seeing-limited observations, is a primary impediment. Due to the variable coverage and depth across our survey footprint, we adopt two different sets of selection criteria for star--galaxy separation and identification of RGB candidates in M64's halo. For the initial analysis that includes the shallower Fields 1 \&\ 2, we adopt a less stringent selection: sources $<$1\farcs5 in size and lying within our visually identified RGB selection region in the $g{-}r$\ vs. $r$\ color--magnitude diagram (CMD). We show a Hess CMD of all sources $<$1\farcs5 in Figure \ref{fig:m64-wide} (top left). For Field 3, and our quantitative analyses of M64's halo, we adopt a stringent selection; to pass, sources must be $<$0\farcs75 in size and must be no further from the $g{-}r$/$r{-}i$\ color locus \citep[Figure \ref{fig:m64-wide}, top center; see also][]{smercina2020} than the quadrature sum of their $g{-}r$\ and $r{-}i$\ uncertainties. RGB candidate stars are then selected in the $g{-}i$\ vs. $i$\ CMD (Figure \ref{fig:m64-wide}, top right). 

In Figure \ref{fig:m64-wide} (bottom) we show a spatial map of RGB star candidates selected using our less stringent criteria for all three fields. We have applied nearest-neighbor filtering to the map, such that only sources with a seventh-nearest neighbor within 40\arcsec\ are displayed, to reduce the presence of the numerous background sources that contaminate this selection.\footnote{Using the \texttt{match\_coordinates\_sky} function in \texttt{astropy}, setting \texttt{nthneighbor}\,=\,7 and selecting only sources with a calculated separation $<$\,40\arcsec. Given the depth of our RGB selection region in $r$-band, this corresponds to an ``effective'' surface brightness of $\mu_r\,{\sim}\,30$\ mag\,arcsec$^{-2}$, assuming a 10\,Gyr, [M/H]\,$=$\,$-$1 stellar isochrone.} Remarkably, a very prominent substructure is visible in M64's southern halo, $\sim$30\textdegree\ counterclockwise from its major axis, extending out $\sim$45\,kpc to a bright shell. On the opposing (northern) side, there is a much more diffuse but similarly shaped plume, which extends out to $>$100\,kpc. These structures are characteristic of a radial merger, where stars follow a plunging orbit \citep[e.g.,][]{johnston2008,amorisco2015,hendel&johnston2015}. The difference in morphology between the northern and southern features suggests that the denser southern shell is the younger structure, with a still-distinct pileup of stars near the apocenter (similar to the simulations of \citealt{cooper2011}). 

Figure \ref{fig:m64-channel} shows a density map of selected RGB stars in all three fields, with color mapped to three bins of inferred photometric metallicity. The metallicity of each star was estimated using a grid of PARSEC isochrones \citep{bressan2012}, with age of 10\,Gyr and metallicities ranging from [M/H]\,=\,$-$2 to 0 in steps of $\Delta$[M/H]\,=\,0.05\,dex. Visually, the southern shell and northern complement are reasonably high metallicity --- [M/H]\,$>$\,$-$1.2. There does not appear to be a clear radial gradient in metallicity.

\section{Properties of M64's Halo and its Progenitor}

In this section, we explore the properties of M64's radial merger remnant and its probable progenitor, in greater detail. First, using existing HST data, we show that our Subaru HSC observations are probing accreted stars in M64's halo, with little contamination from the disk. We then estimate the mass and metallicity of the southern shell feature observed in Field 3. We use this insight to estimate the initial properties of the disrupted progenitor dwarf galaxy and discuss the results in the context of other nearby systems.

\subsection{Mass and Metallicity of the Southern Shell}
\label{sec:sshell}
\begin{figure*}[t]
    \centering
    \includegraphics[width={\linewidth}]{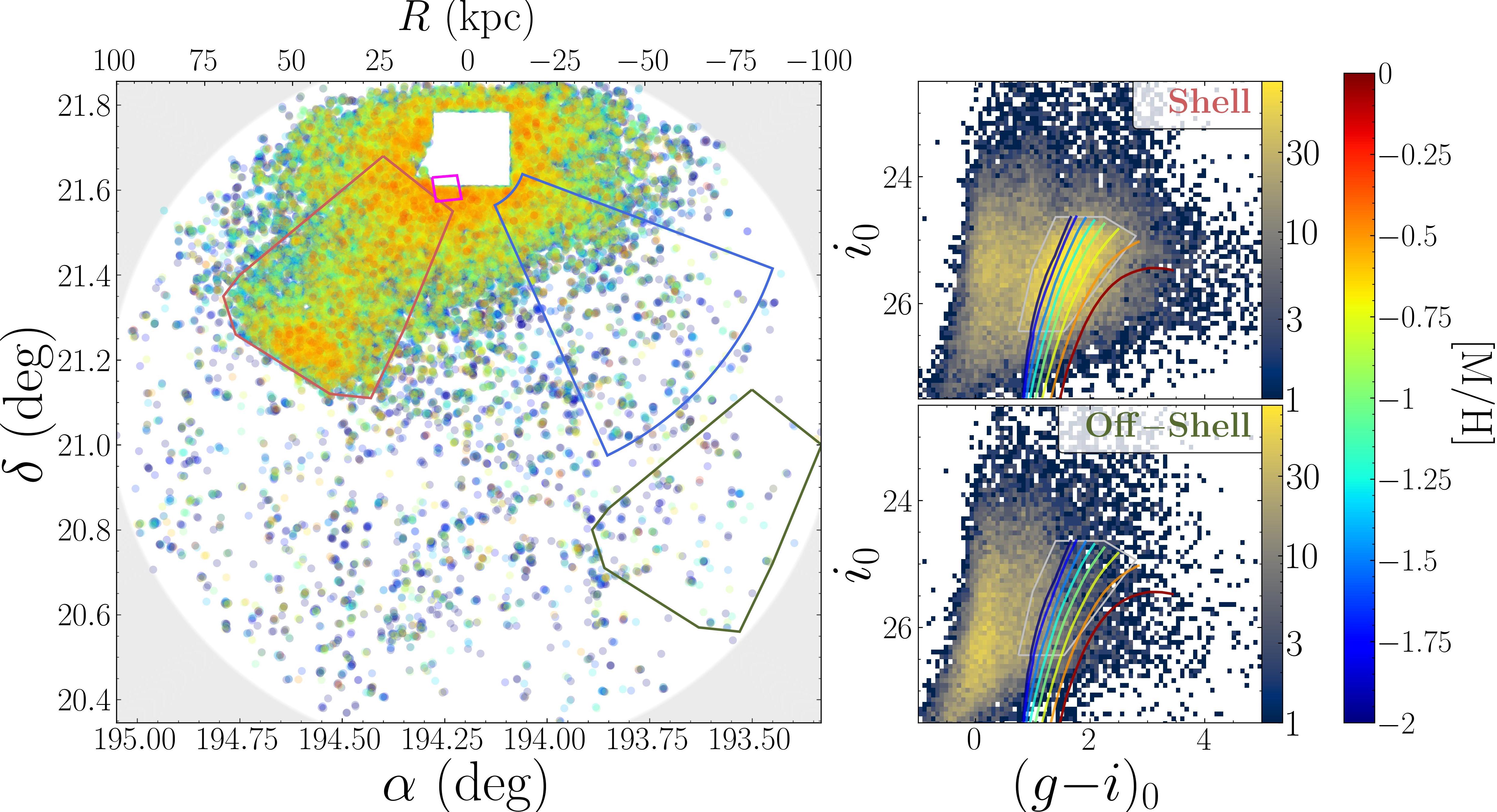}
    \caption{\textit{Left}: Map of high-quality RGB stars in Field 3, color coded by photometrically inferred metallicity. Nearest-neighbor filtering has been applied to reduce background contamination, as in Figure \ref{fig:m64-wide}. The chosen shell selection region is shown in red, with the equivalent background region in olive green. The minor axis region is shown in blue and the archival ACS field from PHANGS-HST in magenta. \textit{Right}: $g{-}i$\ vs.\ $i$\ CMDs of high-quality stellar candidates in the shell (top) and off-shell background (bottom) regions. Stellar isochrones with age 10\,Gyr and metallicities from $-$2 to 0, with 0.25\,dex spacing, are overlaid on both. A strong RGB is visible in the shell region, as expected, but is not visible in the background region.}
    \label{fig:m64_shell-cmds}
\end{figure*}

In Figure \ref{fig:m64_shell-cmds} we show the map of high-quality RGB stars, color coded by inferred photometric metallicity from our grid of PARSEC isochrones. We choose a selection region for the southern shell that encompasses all visibly structured debris (beginning at $\sim$10\,kpc). We show this selection region overlaid on the RGB star map in Figure \ref{fig:m64_shell-cmds} (left), as well as the CMD for all good-quality stars within (top right). We estimate an average metallicity of [M/H]\,$=$\,$-$1 for this shell region. 

To estimate the mass, we first choose a region of identical 0.12\,deg$^2$\ area at the edge of Field 3 ($\sim$100\,kpc from M64) to estimate the background. We show the CMD of high-quality stars in this background region in Figure \ref{fig:m64_shell-cmds} (bottom right), to confirm that there is no visible RGB. Next, using the results of the ASTs (see \S\,\ref{sec:obs}), we estimate the total completeness using the stringent stellar selection within the shell region and correct the background-subtracted counts to a bulk number of expected RGB stars in the shell region. Last, using a [M/H]\,$=$\,$-$1 isochrone (with a Kroupa IMF and 10\,Gyr age), we convert these RGB counts to an estimated total stellar mass in the southern shell of $M_{\rm \star,shell}\,{=}\,1.80\,{\pm}\,0.54{\times}10^8\ M_{\odot}$. Following past works \citep[e.g.,][]{harmsen2017}, we adopt a 30\% uncertainty on all star-count and isochrone-derived stellar masses.\footnote{Note that uncertainties on stellar masses inferred from star counts are dominated by uncertainty on the properties of the applied stellar evolutionary model, mainly that precise stellar ages and metallicities cannot be inferred from our photometry alone. For a reasonable range of possible ages and metallicities for an ancient stellar halo population, this systematic uncertainty impacts stellar mass inference at a $\sim$\,30\% level.}

An important consideration for this analysis is that it has been well documented that radial mergers of the kind M64 is experiencing may pull \textit{in situ} (i.e.\ disk) stars from the central during pericenter passages \citep[e.g.,][]{gomez2017,laporte2018,sanderson2018}, complicating the interpretation of visible debris. To investigate this possibility in M64, we use archival HST observations from the PHANGS-HST survey \citep{lee2022}. The CMD of stars detected in this single Advanced Camera for Surveys (ACS) field in M64's outer disk is shown in Figure \ref{fig:m64-hst} (top left). As discovered by \cite{kang2020}, M64's outer disk exhibits two distinct stellar populations: a metal-rich population, described well by a stellar isochrone with [M/H]\,$=$\,$-$0.3, and a metal-poor population described well by a stellar isochrone with [M/H]\,$=$\,$-$1. 

We convert the PHANGS-HST photometry to the SDSS $g$/$i$\ filters and ``degrade'' it to the Subaru HSC depth achieved in Field 3, using $g$- and $i$-band completeness functions (shown in Figure \ref{fig:m64-hst}, right) derived from our ASTs. We show the predicted CMD of these converted HST sources in Figure \ref{fig:m64-hst} (top, second from left), including predicted detections (black) and nondetections (gray) at the depth of Field 3. The metal-rich disk population, visible in the HST CMD, is largely invisible at the depth of our Subaru observations, due to the depth of the $g$-band observations and the red colors of RGB stars. The HST-identified disk stars are sufficiently red that they fall faintwards of our g-band limit, even at the tip of the RGB (TRGB), while the metal-poor halo population is well-recovered by our RGB selection region. We also show an observed CMD of detected Subaru sources in the same region as the HST observations (Figure \ref{fig:m64-hst}; top, third from left), which matches the predicted Subaru-depth HST CMD well. 

As final evidence that our Subaru observations are purely probing the accreted stars belonging to the shell progenitor, we compare radial mass density profiles derived from the metal-poor HST stars, metal-rich HST stars, and Subaru stars (Figure \ref{fig:m64-hst}, bottom). For each of the HST populations (selected from the regions shown in Figure \ref{fig:m64-hst}, top left), we calculate their densities in 0.5\,kpc bins. As expected \citep[e.g.,][]{kang2020}, the disk and halo populations exhibit very different profiles, with the disk following a steep profile such as observed by \cite{watkins2016} (shown in Figure \ref{fig:m64-hst}, gray squares), and the halo stars following a shallower power law profile. For the Subaru profile, we calculated the density in 1\,kpc bins for radii $<$25\,kpc, and 1.5\,kpc bins $>$25\,kpc. We then background-subtracted and converted counts to stellar mass in each bin, following the method described above. We used the ASTs to calculate completeness and correct counts separately for each radial bin. The RGB star-count-derived Subaru density profile matches the HST halo profile almost exactly and is distinct from the steeper profile of the metal-rich disk population. 

In summary, the estimated stellar mass of $1.80{\times}10^8\ M_{\odot}$\ within the shell appears to be entirely composed of stars accreted from the progenitor satellite.

\subsection{Total Mass of the Progenitor}
\label{sec:progenitor}

In the previous two sections, we presented the discovery of a spectacular radial merger remnant in M64's stellar halo (\S\,\ref{sec:results}) and estimated the mass and metallicity of the southern shell feature in our Field 3 with three-filter photometry (\S\,\ref{sec:sshell}). In this section, we estimate the total stellar mass of the progenitor galaxy, which includes the northern plume visible in our shallower Fields 1 \&\ 2 and the presence of accreted debris in M64's inner disk. We also account for possible contribution from M64's more diffuse stellar halo. 

\begin{figure*}[t]
    \centering
    \includegraphics[width={0.94\linewidth}]{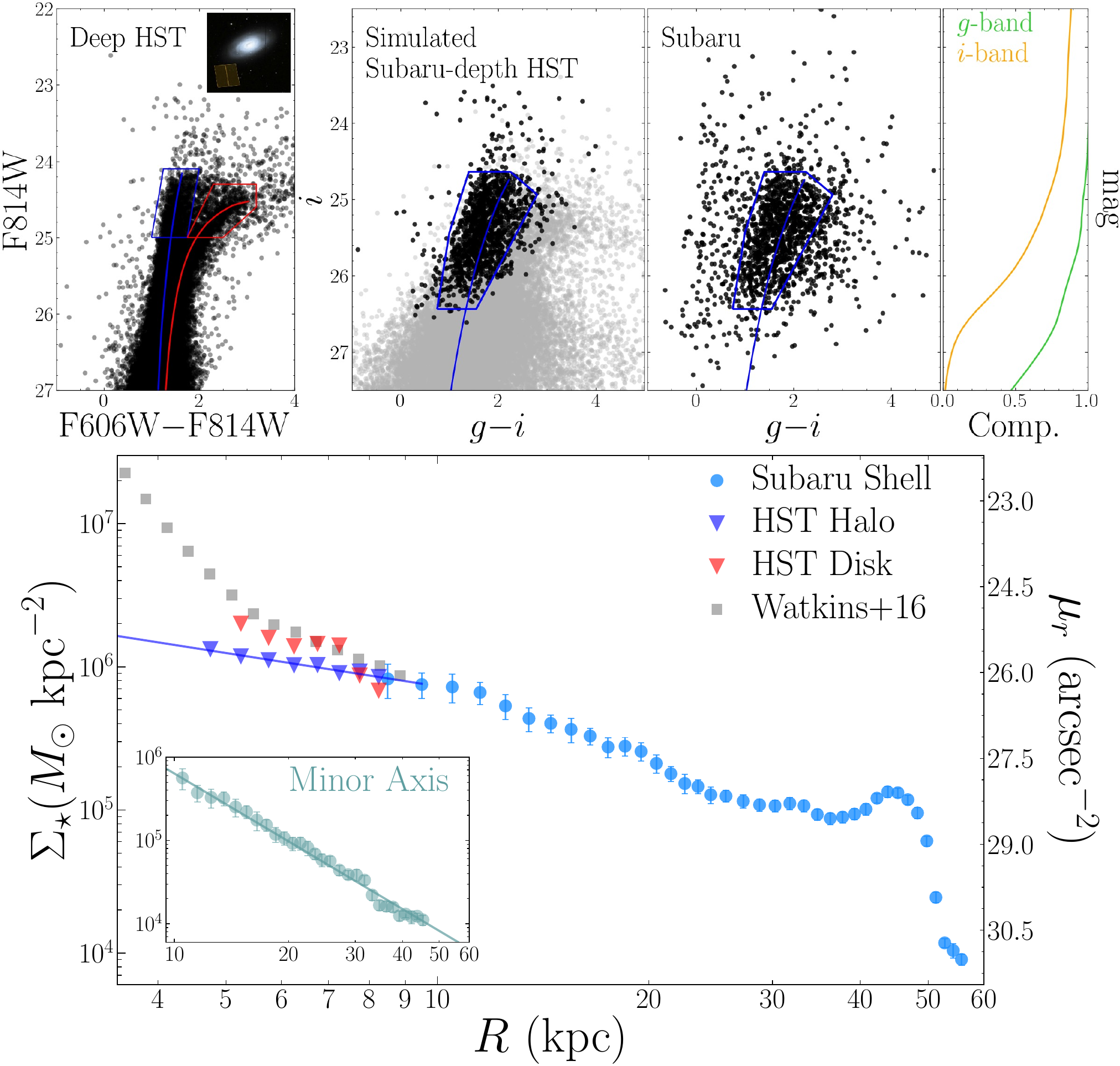}
    \caption{\textit{Top left}: HST photometry from a single ACS field on the outer edge of M64's disk \citep[from PHANGS-HST;][]{lee2022}. Two distinct populations are visible: the metal-rich disk (red) and a metal-poor accreted component (blue). Best-fit isochrones of [M/H]\,$=$\,$-$1 and $-$0.3 are overlaid. \textit{Top, second from left}: HST photometry converted to $g$\ and $i$\ filters and degraded to the depth of the Subaru HSC observations. The Subaru RGB selection region and a [M/H]\,$=$\,$-$1 isochrone model are shown in blue. Sources predicted to be ``detected'' after applying AST-derived $g$\ and $i$\ completeness functions are shown as black circles, while sources predicted to be ``undetected'' are shown in gray. M64's metal-poor halo population is well recovered within our RGB selection region, while the metal-rich disk population is predicted to be largely invisible. \textit{Top, third from left}: Good stellar candidates from Subaru, within the archival ACS field. The RGB is remarkably similar to the predicted CMD of the HST photometry at our Subaru depth, suggesting that our RGB selection criteria are cleanly selecting M64's halo without substantial contamination from the metal-rich disk. \textit{Top right}: $g$\ and $i$\ completeness functions for stringently selected RGB candidates in our Subaru HSC observations, derived from ASTs and applied to the HST photometry. \textit{Bottom}: Stellar mass surface density profiles calculated from Subaru RGB counts along the stream (light blue circles), HST metal-poor RGB counts (blue triangles), and HST metal-rich RGB counts (red triangles). A power law fit to the accreted HST stars is shown. The Subaru minor axis profile and best power law fit are shown in the inset panel. Corresponding $r$-band surface brightness, calculated from the same 10\,Gyr isochrone used for stellar mass estimation, is shown on the right-hand axis. We also show the integrated light minor axis profile measured by \cite{watkins2016} for reference (converted using a $M/L_V$\ ratio from \citealt{bell&dejong2001} and a $b/a$\ of 0.57). The profile derived from RGB stars detected with Subaru, in which the shell is clearly visible, matches well the metal-poor halo profile from HST.}
    \label{fig:m64-hst}
\end{figure*}

Estimating the mass of the northern plume is more complicated than the southern shell as our imaging in the northern fields is both shallower and lacking the critical third filter, $i$-band, resulting in substantially higher contamination of selected RGB stars. With this limitation in mind, to estimate the total stellar mass of the northern plume, we first measure the number of loosely selected RGB stars in M64's southern shell region and subtract off the background counts using the same region shown in Figure \ref{fig:m64_shell-cmds}. We then measure the number of RGB stars in the northern plume (visually identified), subtract off the estimated background (using the less stringent RGB selection) in the Field 3 region shown in Figure \ref{fig:m64_shell-cmds} to be dominated by background sources, and, taking the ratio of loosely selected RGB stars in the northern plume and southern shell, scale the number of RGB stars in the southern shell to the  the northern plume in order to estimate the mass --- resulting in a very similar mass of $M_{\rm\star,plume}\,{=}\,1.42\,{\pm}\,0.71{\times}10^8\ M_{\odot}$. As the mass of the northern plume is estimated in a less direct way, we conservatively adopt a higher uncertainty on this mass. We estimate that the contrast of the northern plume against its respective background is $\sim$3$\times$\ lower than the southern shell and therefore assume a factor of $\sqrt{3}$\ larger error on the mass of the plume, or $\sim$50\%.

The presence of only two opposing shell- or plume-like features, and the visual difference in density between them, suggests that the merger responsible for forming M64's distinct halo is ongoing. Building intuition from galaxy formation simulations, the northern, more diffuse plume likely formed on an earlier passage of the infalling satellite, from more loosely bound stars, while the denser southern shell likely represents a pileup of stars near apocenter following a more recent passage, containing more bound stars from the  inner regions of the disrupting satellite \citep[e.g.,][]{quinn1984,johnston2008,cooper2011}. To support this, in the right two panels of Figure \ref{fig:m64-channel} we compare the morphology of M64's halo to the Aquarius AqF2 simulation of \cite{cooper2011}. Across the two snapshots, separated by 2\,Gyr, the AqF2 simulation shows the progression of a minor merger between the central ($M_{\star}\,{=}\,1.3{\times}10^{10}\,M_{\odot}$) and a satellite ($M_{\star}\,{=}\,7{\times}10^6\,M_{\odot}$) on a radial orbit. Roughly 2\,Gyr after the first pericenter passage, the debris field looks suprisingly similar to the morphology of M64's halo, including the dense southern shell and diffuse northern plume. There are important differences between this simulation and M64, mainly that the progenitor satellite is substantially less massive (by $\sim$2 orders of magnitude in stellar mass) and the resulting debris field is $\sim$3$\times$\ more compact than in M64. While a more detailed simulation is needed to explain the properties of M64's halo exactly, the qualitative similarities in this comparison support the conclusion that M64' halo originates from a single progenitor, which was accreted relatively recently from a satellite with a largely radial initial orbit.

The morphology of the shell and plume suggests that a large fraction of the progenitor's stellar mass should be contained in these structures. However, we expect a large amount of the progenitor's stellar mass to also lie near the center of M64. Indeed, we observe this accreted population at radii $<$10\,kpc in the HST observations \citep[Figure \ref{fig:m64-hst} and][]{kang2020}. To estimate the contribution from this inner accreted material, converting to stellar mass density as described above in \S\,\ref{sec:sshell}, we fit a power law to the radial density profile from the metal-poor HST-selected RGB stars and find that they are well described by a power law of slope $-$0.7. We integrate this profile from the galaxy center out to the start of the shell substructure at 10\,kpc, finding a total accreted stellar mass within the inner disk of $2.13\,{\pm}\,0.64{\times}10^8\ M_{\odot}$. Combining this with our inference of the mass present in the visible shell and plume structures, we therefore estimate a total stellar mass of the disrupted progenitor of $5.35\,{\pm}\,1.1{\times}10^{8}\ M_{\odot}$\ --- a merger ratio of $\sim$52:1 relative to M64's stellar mass. This is very similar to the stellar mass of the SMC\footnote{While no detailed estimates of the SMC's stellar mass have been published, we estimate a stellar mass of $\sim$4--5$\times$10$^8\ M_{\odot}$, assuming the $K$-band luminosity of 7.1$\times$10$^8\ L_{\odot}$\ reported in \cite{karachentsev2013} and a $M/L_K$\ ratio of $\sim$0.6 \citep{bell&dejong2001}.} and with an estimated metallicity of [M/H]\,$\sim$\,$-$1, would make the progenitor completely consistent with the stellar mass--metallicity relation \citep{kirby2013}. Incidentally, the total accreted stellar mass of $2.13{\times}10^8\,M_{\odot}$\ that we infer resides within M64's disk ($R\,{<}\,10$\,kpc) is nearly identical to the upper limit placed by \cite{rix1995} on the amount of possible counter-rotating stellar material in M64's disk, based on integrated stellar kinematics. 

There is additional material in the halo outside of the visible substructure, though it is unclear if this stellar halo material is associated with the recent merger. It does appear to have a similar metallicity of [M/H]\,${\sim}$\,$-$1 (Figure \ref{fig:m64_shell-cmds}). The stellar populations along the minor axes of galaxies have been shown to be excellent tracers of diffuse stellar halo properties \citep[e.g.,][]{monachesi2016b,harmsen2017,dsouza&bell2018a}, and in this case M64's minor axis has the added advantage of being largely perpendicular to the shell structure. While we have already estimated the accreted material in the inner disk, we can therefore estimate the amount of stellar mass in this diffuse outer halo from the minor axis profile. We show the minor axis profile in Figure \ref{fig:m64-hst} (inset panel), calculated identically to the shell profile shown in the same figure. M64's minor axis profile is well described by a power law of slope $-$2.7. Following \cite{harmsen2017} we integrate this power law profile from 10--40\,kpc, estimating a total stellar mass $2{\times}10^8\,M_{\odot}$\ in the outer halo. If M64's diffuse halo is connected to its recent merger, then this gives an upper limit on the progenitor stellar mass of ${\lesssim}7{\times}10^8\,M_{\odot}$\ --- still consistent with our inference that the progenitor was an SMC-mass galaxy. 

In summary, the stellar populations in M64's disk, their kinematics, and the structure and abundances of populations in its stellar halo are consistent with its accretion of a gas-rich, SMC-like progenitor in the recent past. 

\section{Discussion and Conclusions}
\label{sec:conclusions}
In this Letter, we have presented a first-ever view of the resolved stellar populations in the stellar halo of M64, using Subaru HSC. Detected RGB stars in M64's halo reveal a spectacular ongoing radial merger, including a diffuse northern plume and a dense southern shell. We estimate a stellar mass in the southern shell of $M_{\rm\star,shell}\,{=}$\,1.80\,${\pm}\,0.54\,{\times}$10$^8\ M_{\odot}$, with an average metallicity of [M/H]\,$=$\,$-$1. Using existing HST observations, which separate M64's metal-rich disk and metal-poor accreted populations, we determine that the stars detected in our Subaru survey are entirely consistent with the accreted population, with little \textit{in situ} contamination from the disk. Scaling from the mass of the southern shell, we estimate a nearly equal stellar mass in the northern plume of $M_{\rm\star,plume}\,{=}$\,1.42\,${\pm}\,0.71\,{\times}$10$^8\ M_{\odot}$. Using the profile of inferred accreted stars in M64's inner disk, from existing HST observations, we estimate that another $2.13\,{\pm}\,0.64{\times}10^{8}\,M_{\odot}$\ of accreted stellar mass exists in M64's interior, giving a total stellar mass of the progenitor of $M_{\rm\star,prog}\,{\simeq}$\,5$\times$10$^8\ M_{\odot}$. 

This progenitor satellite was, therefore, likely remarkably similar to the SMC in both stellar mass and metallicity. This is important for understanding the origins of M64's counter-rotating gas disk, as its accreted gas component is very similar in mass ($\simeq$5$\times$10$^8\ M_{\odot}$; \citealt{walter2008}) to the gas content of the SMC ($\simeq$4$\times$10$^8\ M_{\odot}$; \citealt{dickey2000}). The morphology of the debris implies that the disrupting satellite experienced a last pericenter passage likely $<$1\,Gyr ago \citep[e.g.,][]{johnston2008,cooper2011}, and the merger has been ongoing for the last several Gyrs. This means that any flyby interaction between M64 and the distant Coma P \citep{brunker2019} would have occurred during M64's merger with the current progenitor. We therefore assert that the shell progenitor is the more likely donor of M64's counter-rotating gas as it is much more massive (though as noted by \citealt{brunker2019}, Coma P is surprisingly gas rich). This makes sense in the visual context of M64's halo: the plume--shell structure suggests that the satellite's last pericenter passage occurred from northwest to southeast, which is counter to the rotational direction of M64's stars, at least in projection. 

The likely origins of the Evil Eye's counter-rotating gas are at last revealed, after decades of curiosity. We suggest that its outer gas disk was accreted recently during a 52:1 merger with an SMC-mass galaxy and is now colliding with an existing inner gas disk, fueling a burst of star formation at the disk--disk interface and driving the visible dust lanes from which it earns its name. Future observational and theoretical studies will help test this idea. Resolved spectroscopic studies of the kinematics and abundance gradients of stars across the debris field, enabled by upcoming wide-field integral field unit instruments, would provide definitive evidence that the structures were accreted from the same progenitor. Moreover, studies using hydrodynamic simulations of similar galaxies in the Illustris and EAGLE suites have begun to explore the phenomenon of kinematically misaligned gas and stars \citep[e.g.,][]{starkenburg2019,khoperskov2021,lu2021,casanueva2022}. Results from these simulations may help inform the possible connection between M64's counter-rotating gas and the evolution of its recent merger. For example, it is currently unclear when the gas would have been accreted in our presented scenario.

While a fascinating case study of how mergers shape and drive the properties of nearby galaxies, M64's unique set of properties make it an exciting target for understanding stellar halos generally. Connecting the counter-rotating gas to a radial merger event would place a strong observational prior on the satellite's orbital properties, mainly that it must have been partially retrograde. Such a reliable prior on a halo progenitor's orbital properties has only been achieved for the MW \citep[e.g.,][which has a very different merger history]{helmi2018} and otherwise requires model-dependent assumptions \citep[e.g.,][for M31]{hammer2018}. Efforts to reproduce M64's halo properties in models and galaxy formation simulations therefore has the potential to contribute more broadly to a better understanding of the buildup of stellar halos around galaxies. \\

A.S.\ was supported by NASA through grant No.\,GO-14610 from the Space Telescope Science Institute, which is operated by AURA, Inc., under NASA contract NAS 5-26555. E.F.B.\ was partly supported by the National Science Foundation through grant 2007065 and by the WFIRST Infrared Nearby Galaxies Survey (WINGS) collaboration through NASA grant NNG16PJ28C through subcontract from the University of Washington. A.M. gratefully acknowledges support by the ANID BASAL project FB210003, FONDECYT Regular grant 1212046, and funding from the Max Planck Society through a “PartnerGroup” grant.

Based on observations utilizing Pan-STARRS1 Survey. The Pan-STARRS1 Surveys (PS1) and the PS1 public science archive have been made possible through contributions by the Institute for Astronomy, the University of Hawaii, the Pan-STARRS Project Office, the Max Planck Society and its participating institutes, the Max Planck Institute for Astronomy, Heidelberg and the Max Planck Institute for Extraterrestrial Physics, Garching, The Johns Hopkins University, Durham University, the University of Edinburgh, the Queen's University Belfast, the Harvard-Smithsonian Center for Astrophysics, the Las Cumbres Observatory Global Telescope Network Incorporated, the National Central University of Taiwan, the Space Telescope Science Institute, the National Aeronautics and Space Administration under grant No. NNX08AR22G issued through the Planetary Science Division of the NASA Science Mission Directorate, the National Science Foundation grant No. AST-1238877, the University of Maryland, Eotvos Lorand University (ELTE), the Los Alamos National Laboratory, and the Gordon and Betty Moore Foundation.

Based on observations obtained at the Subaru Observatory, which is operated by the National Astronomical Observatory of Japan, via the Gemini/Subaru Time Exchange Program. We thank the Subaru support staff --- particularly Tsuyoshi Terai, Sakurako Okamoto, and Tom Winegar --- for invaluable help in preparing the observations and navigating the Subaru Queue Mode.

The authors wish to recognize and acknowledge the very significant cultural role and reverence that the summit of Maunakea has always had within the indigenous Hawaiian community. We are most fortunate to have the opportunity to conduct observations from this mountain.

\software{\texttt{HSC Pipeline} \citep{bosch2018}, \texttt{Matplotlib} \citep{matplotlib}, \texttt{NumPy} \citep{numpy-guide,numpy}, \texttt{Astropy} \citep{astropy}, \texttt{SciPy} \citep{scipy}, \texttt{SAOImage DS9} \citep{ds9}}

\bibliographystyle{aasjournal}
\bibliography{references}

\end{document}